\documentclass[conference]{IEEEtran}
\IEEEoverridecommandlockouts

\usepackage{cite}
\usepackage{amsmath,amssymb,amsfonts}
\usepackage{algorithmic}
\usepackage{graphicx}
\usepackage{textcomp}
\usepackage{xcolor}
\usepackage{verbatim}
\usepackage{comment}
\usepackage{booktabs, multirow, makecell, xcolor, adjustbox}
\definecolor{rowgray}{gray}{0.95}
\usepackage[normalem]{ulem}
\def\BibTeX{{\rm B\kern-.05em{\sc i\kern-.025em b}\kern-.08em
    T\kern-.1667em\lower.7ex\hbox{E}\kern-.125emX}}

\usepackage{eso-pic}
\AddToShipoutPictureFG*{\AtPageUpperLeft{\put(36,-20){\parbox[t]{540pt}{\fontsize{8}{9}\selectfont
\textcopyright\ 2026 IEEE. Permission from IEEE must be obtained for all other uses, in any current or future media, including reprinting/republishing this material for advertising or promotional purposes, creating new collective works, for resale or redistribution to servers or lists, or reuse of any copyrighted component of this work in other works.\\
Author accepted manuscript. Accepted for publication in the 2026 IEEE Biomedical Circuits and Systems Conference (BioCAS).
}}}}

\begin{document}

\title{\vspace{14pt}An Open-Architecture Dual-Core STM32-Based Neural Interface with Low-Latency On-Device \\AI for Neuromodulation Applications}

\author{\IEEEauthorblockN
        {Sipeng~Liang\IEEEauthorrefmark{1},
        Kun~Cheng\IEEEauthorrefmark{1},
        Xiao~Wu\IEEEauthorrefmark{1},
        Lyu~Wang\IEEEauthorrefmark{1},
        Savanna~Blade\IEEEauthorrefmark{1},
        Jiaxi~Wei\IEEEauthorrefmark{1},
        Ruoheng~Wang\IEEEauthorrefmark{1},\\
        Zhecheng Yin\IEEEauthorrefmark{1},
        Johnson~Song\IEEEauthorrefmark{1},
        Tristan~J.~Lawson\IEEEauthorrefmark{2},
        Alisha~Baldwin\IEEEauthorrefmark{2},
        Muhammad~Shan~Sohail\IEEEauthorrefmark{2}, \\
        Luka~Milosevic\IEEEauthorrefmark{2},
        and~Xilin~Liu\IEEEauthorrefmark{1}
        }
\IEEEauthorblockA{\IEEEauthorrefmark{1}Department of Electrical and Computer Engineering (ECE), University of Toronto, Toronto, ON, Canada\\ \IEEEauthorrefmark{2}Krembil Brain Institute, Toronto Western Hospital, University Health Network (UHN), Toronto, ON, Canada}}
\maketitle

\begin{abstract}

We present an open-architecture bidirectional neural interface for closed-loop neuromodulation built around a dual-core STM32H747 microcontroller and commercial Intan recording and stimulation front-ends. The system supports up to 128 recording channels at 20\,kSps and 64 stimulation channels while enabling low-latency on-device AI inference through a dual-core processing architecture and a two-stream memory hierarchy. To validate the machine learning pipeline, we demonstrate two neural decoding models: a compact multilayer perceptron executing entirely from on-chip memory and a larger temporal neural decoder utilizing external memory staging. Both models successfully decode fingertip velocity from neural spiking activity. As a proof of concept for closed-loop operation, we implement spike-triggered stimulation and achieve an end-to-end latency of 189 $\pm$ 14\,$\mu$s, representing an improvement of three orders of magnitude over a host-PC-mediated closed-loop pathway. These results demonstrate that resource-constrained embedded platforms can support high-channel-count neural interfacing, flexible on-device AI, and sub-millisecond closed-loop neuromodulation.

\end{abstract}


\section{Introduction}
\label{sec:introduction}

Recent advances in AI-based biomarker detection~\cite{liu2021edge} and multimodal home-based rehabilitation monitoring~\cite{yang2024multimodal} have intensified demand for edge machine learning (ML) algorithms capable of running on resource-constrained neural interfaces, with the goal of enabling more effective closed-loop neuromodulation therapies. Researchers developing and validating these models within embedded closed-loop pipelines face significant implementation challenges. The inherent demands of bidirectional neural interface development, such as precise multichannel acquisition timing and balancing parallel recording and stimulation workloads, must be managed alongside on-device inference. AI models can be computationally intensive and variable in execution time, potentially disrupting acquisition on single-core processors; their memory requirements force tradeoffs between model complexity and on-chip SRAM capacity, while offloading to external memory can introduce unwanted access latencies.

These constraints have shaped existing efforts to integrate ML into closed-loop platforms. Many implementations have relied on application-specific integrated circuits, where inference can be optimized in hardware~\cite{topalovic2023neurostack, rizzuto2025sns}; however, these systems are suited to a fixed model and application. Open-architecture research platforms built around commercially available components~\cite{Siegle_Lopez_Patel_Abramov_Ohayon_Voigts_2017,Newman_Zhang_Cuevas-Lopez_Miller_Honda_vanGoes_Leighton_Carvalho_Lopes_Lakunina_etal._2025} have typically deferred biomarker detection to a host PC without dedicated ML integration, often limiting sub-millisecond closed-loop operation~\cite{Siegle_Lopez_Patel_Abramov_Ohayon_Voigts_2017} and requiring custom ML execution pathways. Other systems reporting on-device closed-loop algorithms have implemented comparatively simple real-time computations, such as band-power detection~\cite{ferrero2025closed} or fixed spiking neural network models~\cite{shupe2021neurochip3}.

\begin{figure}[!t]
    \centering
    \includegraphics[width=0.95\linewidth]{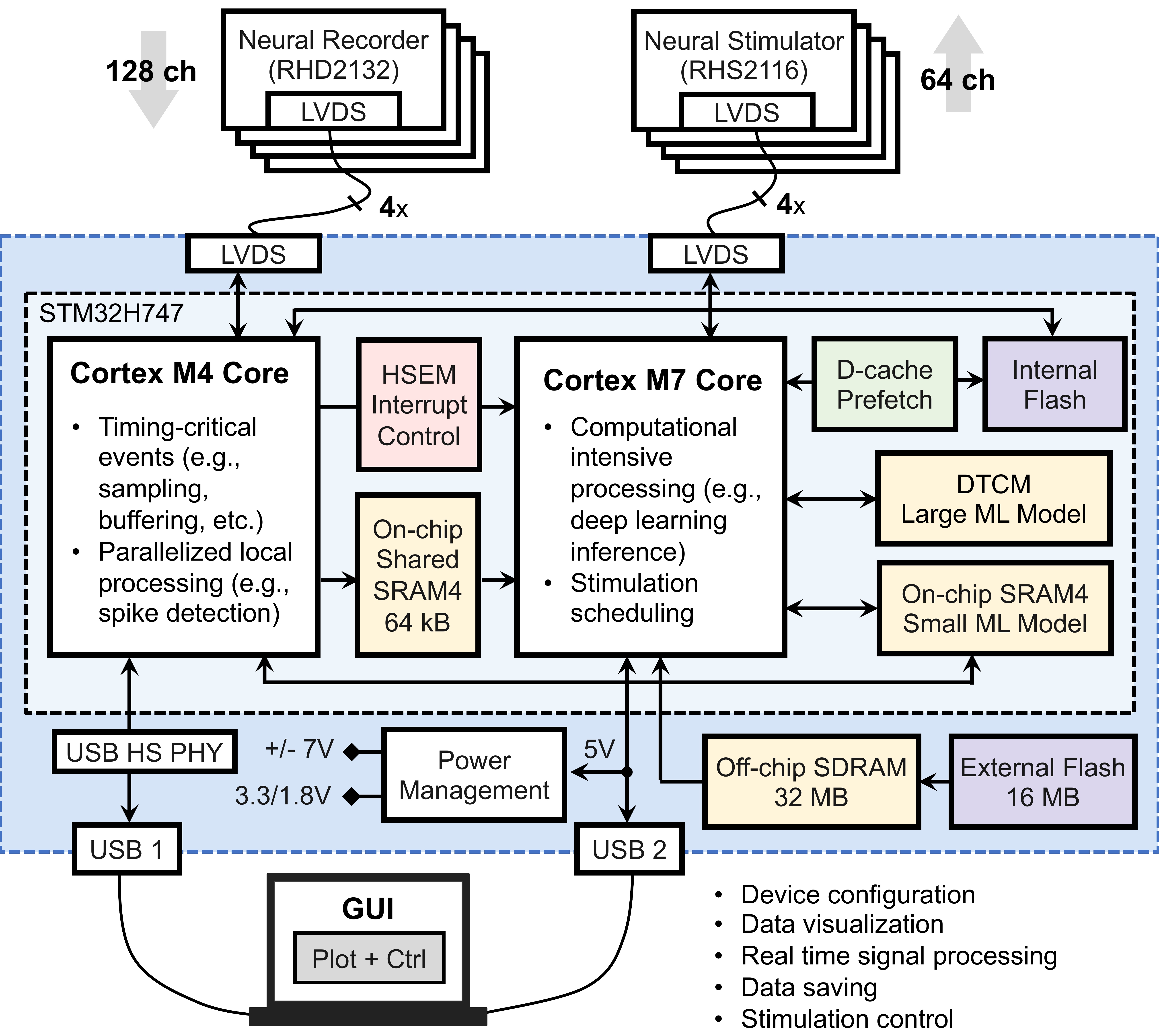}
    \caption{Overview of the STM32H747-based bidirectional neural interface, illustrating the dual-core design and on-device AI pipeline.}
    \label{fig:system_overview} \vspace{-5mm}
\end{figure}

In this work, we present an open-architecture bidirectional neural interface featuring configurable on-device AI. Built around off-the-shelf neural interface modules
(RHD2132 and RHS2116 headstages, Intan Technologies) and a dual-core STM32H747 microcontroller (MCU), the system supports up to 128 recording channels operating at 20\,kSps, as well as 64 stimulation channels. The system partitions time-critical sampling tasks and more variable inference workloads between cores, enabling simultaneous acquisition and inference. The inference pipeline features a two-stream memory architecture, allowing models of various complexities to be iteratively deployed and switched without significant firmware changes.

\section{System Architecture and Hardware Design}\label{hardware}

The key components of the system are an STM32H747 dual-core MCU for orchestrating data acquisition, stimulation, and inference (Fig.~\ref{fig:system_overview}); 4$\times$ RHD2132 modules for recording; 4$\times$ RHS2116 modules for stimulation; and a host PC for configuration and visualization. The STM32H747 MCU was selected to support communication with multiple Intan chips, high-throughput data streaming to the host PC, and computationally intensive ML inference. Each RHD2132 module transmits approximately 10\,Mb/s, requiring its own dedicated SPI channel to ensure continuous real-time data acquisition. The RHS2116 devices are used exclusively for stimulation and can share a single SPI bus without performance degradation, a configuration that also supports synchronous stimulation across multiple RHS devices. High-throughput host communication necessitates the use of a dedicated USB high-speed PHY (USB3300).

\begin{figure}[!b] \vspace{-3mm}
    \centering
    \includegraphics[width=1\linewidth]{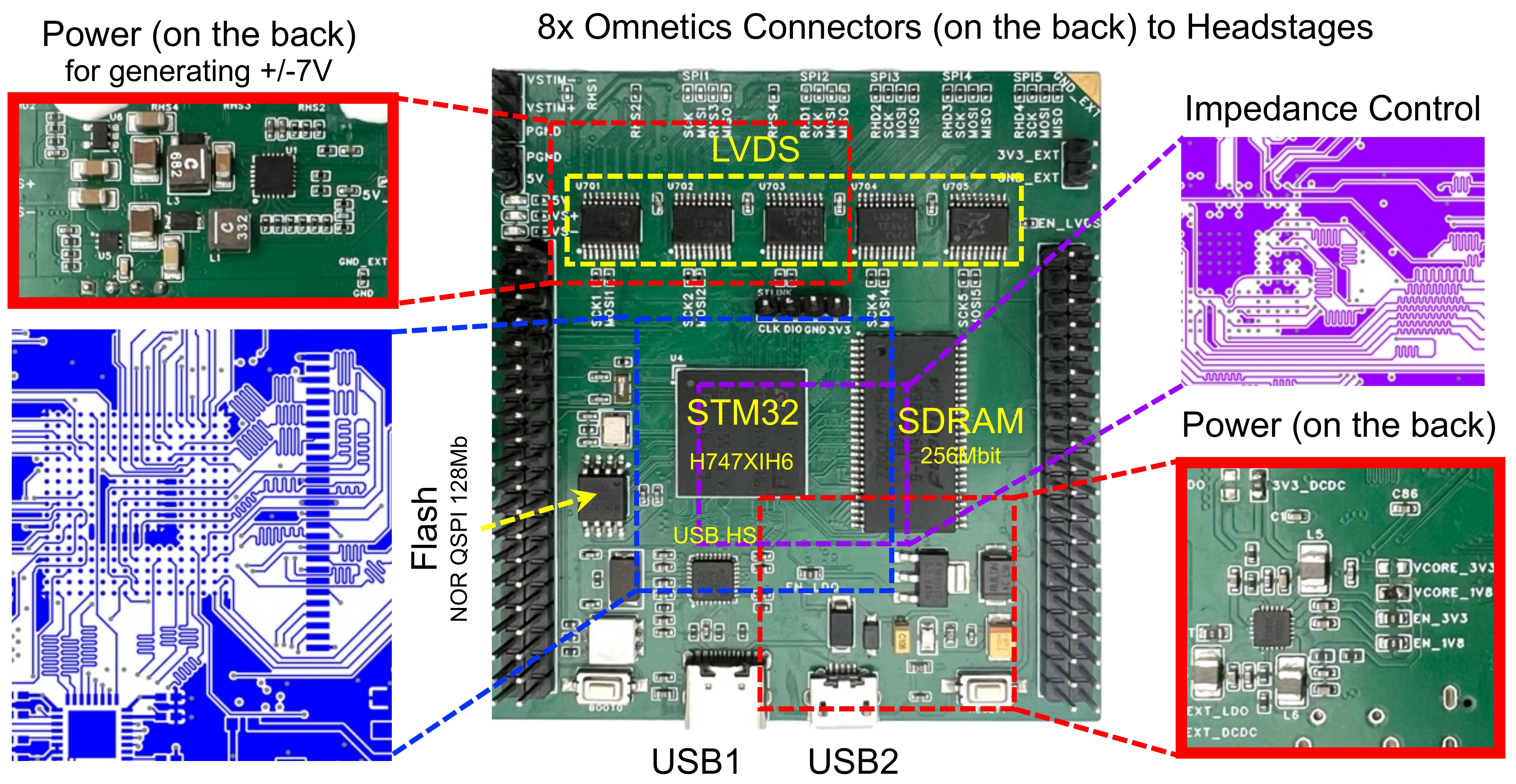} \vspace{-5mm}
    \caption{A photo of the developed PCB with key building blocks highlighted. Special attention was paid to ensure signal integrity for high-speed signals.}
    \label{fig:pcb}
\end{figure}

The STM32H747 MCU meets these requirements through its dual-core ARM Cortex-M7/M4 (M7/M4) architecture (operating at up to 480\,MHz and 240\,MHz, respectively), six SPI peripherals, USB high-speed support with UTMI+ low pin interface (ULPI) for external PHY interfacing, 2\,MB of on-chip Flash, 1\,MB of on-chip SRAM, of which 128\,kB is Data Tightly Coupled RAM (DTCM), and an external memory controller for SDRAM expansion.

A custom PCB (Fig.~\ref{fig:pcb}) was designed to meet three integration requirements: high-speed SPI signal integrity, isolation between analog and digital domains, and power delivery. SPI communication with Intan devices uses low-voltage differential signaling via SN65LVDT41 transceivers to minimize crosstalk and electromagnetic interference. Power is partitioned into four domains: MCU core, digital/peripheral, Intan recording, and RHS stimulation. The $\pm$7\,V stimulation supply is isolated via a DC-DC converter to prevent switching transients from coupling into recording channels.

\section{Firmware Design}\label{software}

The responsibilities of the dual-core STM32H747 MCU are assigned based on the capabilities of each core: the M7 is reserved for computationally intensive inference due to its higher clock speed and L1 cache, while the M4 handles timing-critical data acquisition from the RHD2132 modules. Stimulation control is assigned to the M7 to avoid the additional latency associated with cross-core communication during closed-loop inference-triggered stimulation.
Data acquisition routines are optimized to ensure a stable 20\,kHz sampling rate. Communication between the M4 and the RHD2132 chips is managed by custom lightweight SPI drivers that leverage direct memory access (DMA) to offload data transfer from the CPU. To ensure deterministic sampling at 20\,kHz and avoid timing conflicts with the USB link to the host, the sample-timer, SPI, and DMA-stream interrupts are assigned higher nested vectored interrupt controller (NVIC) priorities than the USB transmit-complete callback, which has inherently non-deterministic latency. Samples are buffered and transmitted in batches to reduce USB transfer routine calls and transmit-complete callbacks.

\subsection{Cross-Core Communication Framework} For tasks requiring cross-core data transfer, a decoupled inter-core communication framework was implemented. The framework consists of a high-throughput, low-latency \textit{fast-path} for inference and a \textit{slow-path} for configuration and telemetry synchronization. Both pathways leverage dedicated shared memory allocated in the 64\,kB D3-domain SRAM4 region (architecturally expandable to high-speed Advanced eXtensible Interface (AXI) SRAM based on throughput demands) and the hardware semaphore peripheral (HSEM) to ensure data integrity and minimize synchronization overhead.

\subsubsection{Fast-Path Communication}
The fast-path is designed for time-critical asynchronous events where data from the M4 must be made visible to the M7. The M4 writes data into a shared SRAM buffer and updates the corresponding control state; when the buffer is ready for consumption, the M4 triggers an HSEM interrupt to wake the M7. The buffer region is cache-disabled to prevent stale-data hazards, allowing for a deterministic handoff between cores. Once the M7 consumes the data and executes inference, it returns to wait-for-interrupt mode, eliminating the need for continuous polling.

\subsubsection{Slow-Path Communication}
The slow-path handles non-time-critical downlink configuration and uplink telemetry. Commands from the GUI reach the M7 via USB and are written to a shared SRAM slot, where the M4 fetches them during periodic housekeeping. Diagnostic logs follow the reverse path: the M4 commits them to a telemetry slot in shared SRAM, which the M7 copies to the USB full-speed transmission buffer during synchronization.

\subsection{Edge Machine Learning Pipeline}
\begin{figure}[!ht]
    \centering
    \includegraphics[width=\linewidth]{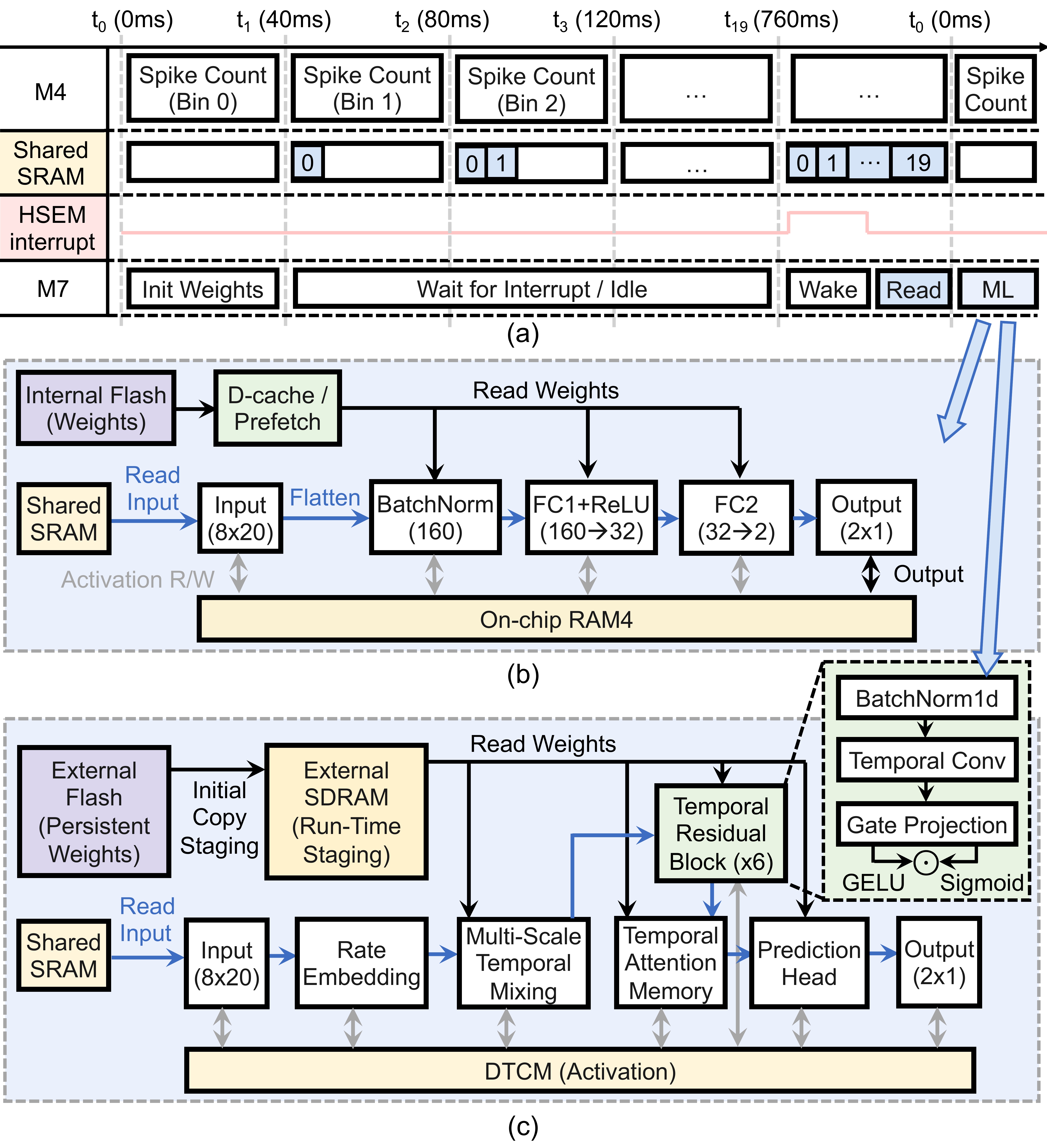} \vspace{-6mm}
    \caption{(a) Timing diagram of the inference pipeline. (b, c) Block diagrams demonstrating the integration of the two characterization models with the on-chip (b) and external (c) memory paths.}
    \label{fig:ml-pipeline}
\end{figure}

The ML pipeline (Fig.~\ref{fig:ml-pipeline}) was developed to support models that take neural spiking activity as input. The M4 performs peak detection via thresholding and bins incoming activity into a sliding 8$\times$20 shared SRAM buffer: 8 channels by 20 temporal bins of 40\,ms, each containing spike counts. Once the buffer is full, the window is handed off to the M7 via the fast-path; inference then proceeds via separate memory paths selected according to model size.

For sufficiently small models, weights are stored in internal Flash and activation buffers are placed in internal RAM. During inference, the M7 reads the weights from internal Flash, using pre-fetch and D-cache to reduce access latency.
For models that exceed the on-chip memory budget, weights are stored in external Flash and copied to external SDRAM during initialization. After initialization, the M7 reads the staged weights from SDRAM as different model stages execute. Activation buffers are placed in DTCM, since these intermediate tensors are frequently accessed during inference.

This two-stream memory access pattern optimizes inference latencies for models across diverse architectures and parameter counts, keeping models with smaller memory footprints entirely on-chip and reducing external memory accesses for larger-footprint models via weight staging.

\section{System Characterization}\label{performance}
This section presents system characterization results validating the acquisition noise floor and inference performance of the two-stream ML pipeline. Additionally, as a proof of concept for closed-loop operation and to establish a lower bound for on-device closed-loop latency, a simple spike-triggered stimulation routine was implemented; its performance is reported.

\subsection{Characterization of the Neural Interface}

\begin{figure}[!ht]
    \centering
    \includegraphics[width=1\linewidth]{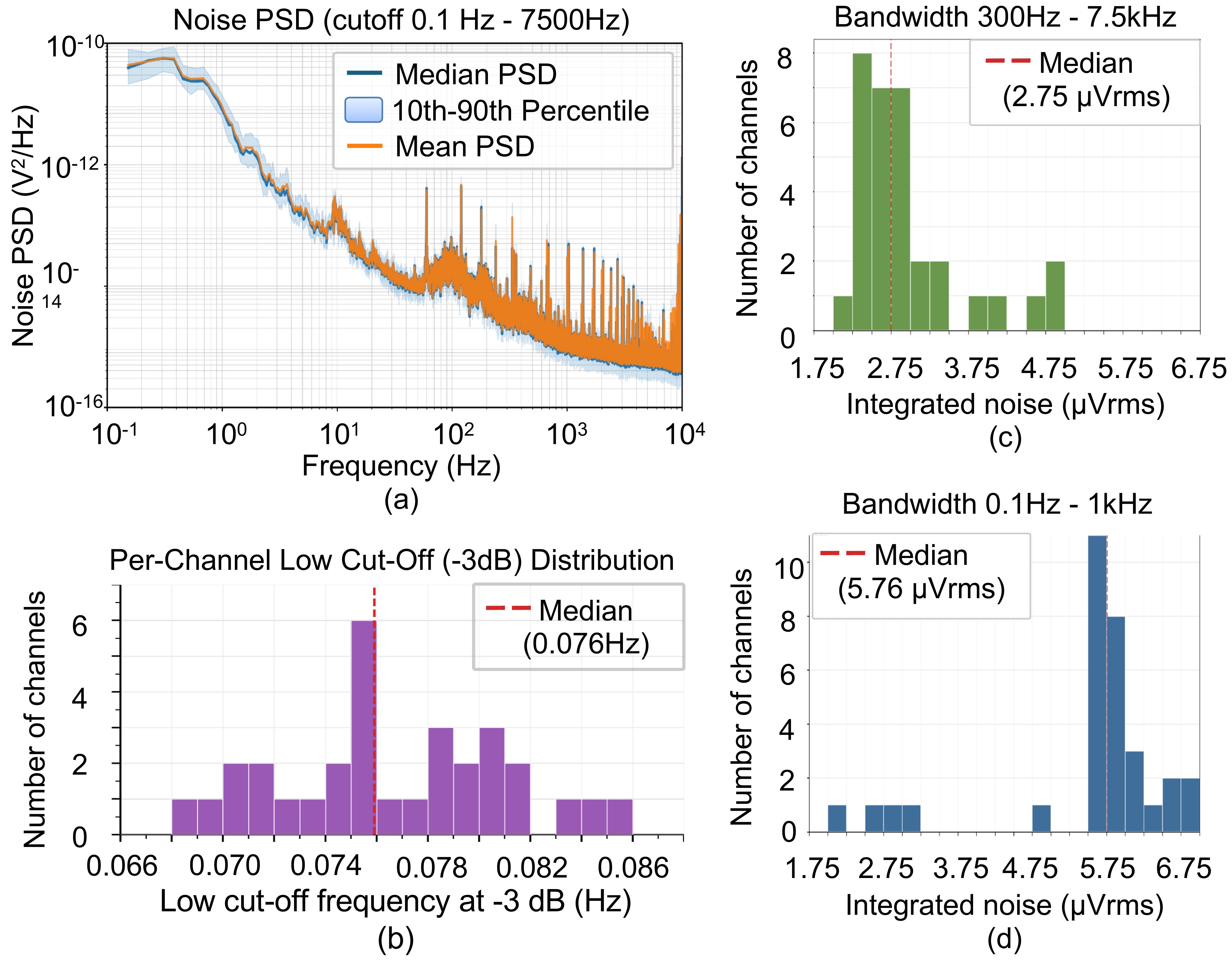}
    \caption{(a) Mean and median PSD of input-referred noise across all channels. (b) Histogram of the measured -3\,dB high-pass cutoff frequencies across channels. (c, d) Histograms of integrated input-referred noise across channels in the 300\,Hz to 7.5\,kHz and 0.1\,Hz to 1\,kHz frequency bands.}
    \label{fig:front-end}
\end{figure}

Fig.~\ref{fig:front-end} characterizes the acquisition performance of the RHD2132 front-end. Input-referred noise (IRN) was measured across all channels; Fig.~\ref{fig:front-end}(a) shows the mean and median noise PSD. IRN was integrated over two pass bands relevant to neural signal acquisition: 300\,Hz to 7.5\,kHz (spiking) and 0.1\,Hz to 1\,kHz (local field potential). Per-channel IRN is represented in Figs.~\ref{fig:front-end}(c, d). The accuracy of the configurable high-pass filter of the RHD2132 was verified by setting the cutoff to 0.1\,Hz and measuring the low-frequency -3\,dB cutoff for each channel (Fig.~\ref{fig:front-end}(b)); the median measured cutoff of 0.076\,Hz confirms the reliability of the filter settings.

\begin{figure}[!ht]
    \centering
    \includegraphics[width=1\linewidth]{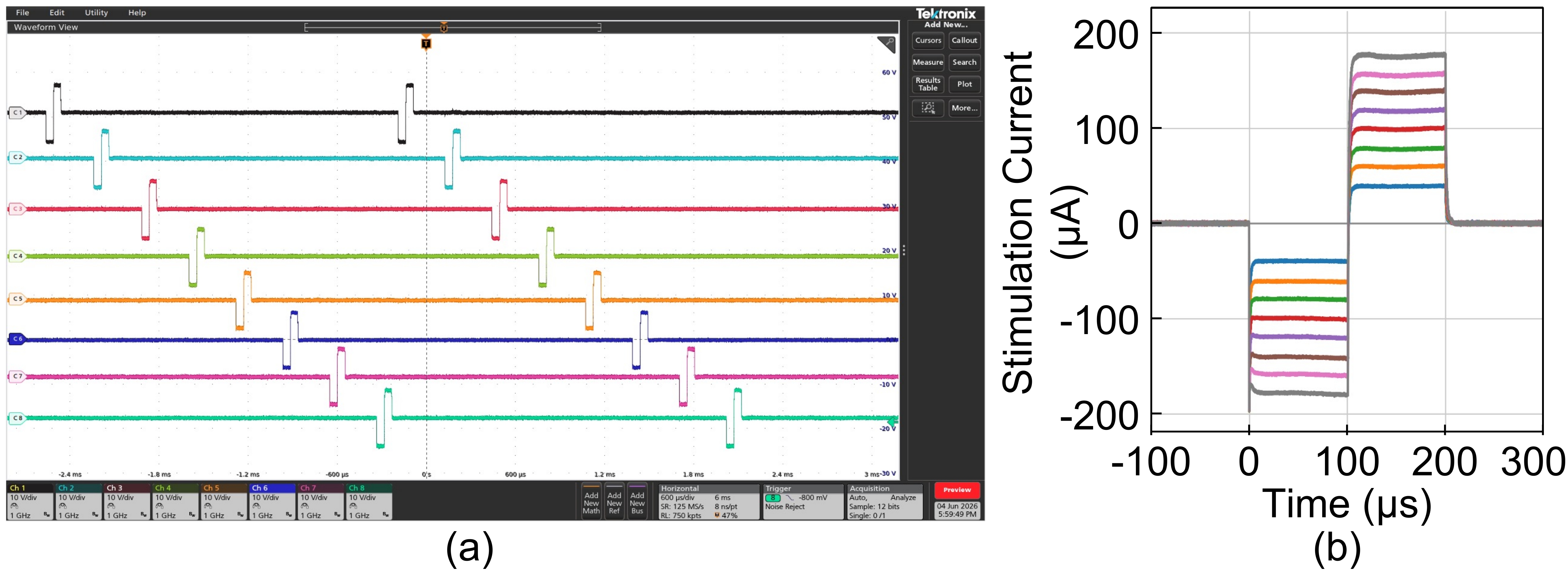} \vspace{-5mm}
    \caption{(a) Representative staggered stimulation from eight channels captured on an oscilloscope.
(b) Overlay of stimulation pulses at different current levels.}
    \label{fig:stimulation}
\end{figure}

Fig.~\ref{fig:stimulation}(a) shows eight-channel stimulation generated by the RHS2116, verifying that the M7 stimulation timing results in balanced biphasic waveforms across varying amplitude settings (Fig.~\ref{fig:stimulation}(b)).
\vspace{-5pt}

\begin{figure*}[!ht]
    \centering
    \includegraphics[width=1\linewidth]{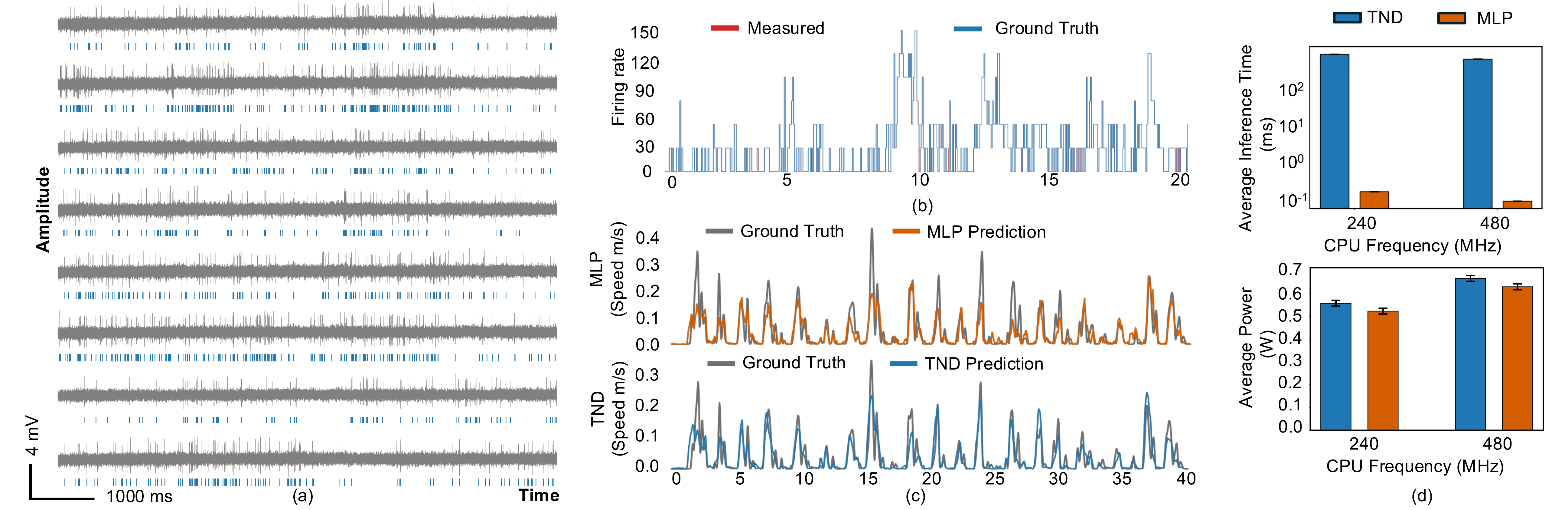}
    \caption{(a) 8-channel on-device spike detection. (b) Detected and ground-truth firing rates for a single channel. (c) Model velocity predictions and ground truth. (d) ML pipeline inference time and power consumption at different CPU frequencies.}
    \label{fig:on-device-AI}
\end{figure*}

\subsection{On-Device Inference}
We evaluated the performance of the inference pipeline using a fingertip-velocity decoding task based on publicly available non-human primate data~\cite{ODoherty_Cardoso_Makin_Sabes_2020}.
Two models were developed for characterization testing. The first is a compact multilayer perceptron (MLP), designed to fit entirely within on-chip memory. The second is a temporal neural decoder (TND) model, designed to evaluate the external-memory execution path. The architectures of both models and their integration with the two memory pathways are documented in detail in Fig.~\ref{fig:ml-pipeline}(b, c).

Fig.~\ref{fig:on-device-AI}(a) demonstrates on-device spike detection using a segment of eight-channel data, with detections represented as spike trains. The accuracy of the on-device firing-rate calculation was assessed against the ground truth, yielding a root-mean-square error of 2.23 (Fig.~\ref{fig:on-device-AI}(b)).
Fig.~\ref{fig:on-device-AI}(c) shows representative segments of the decoded velocity predictions for each model. Both models reproduce the dominant temporal structure of the ground-truth movement signal. The MLP and TND models achieve $R^2$ values of 62.7\% and 64.1\%, respectively.
Finally, Fig.~\ref{fig:on-device-AI}(d) compares the average inference time and MCU power consumption of both models during continuous inference at different CPU frequencies.

\subsection{Closed-Loop Spike-Triggered Stimulation}

\begin{figure}[!ht]
    \centering
    \includegraphics[width=1\linewidth]{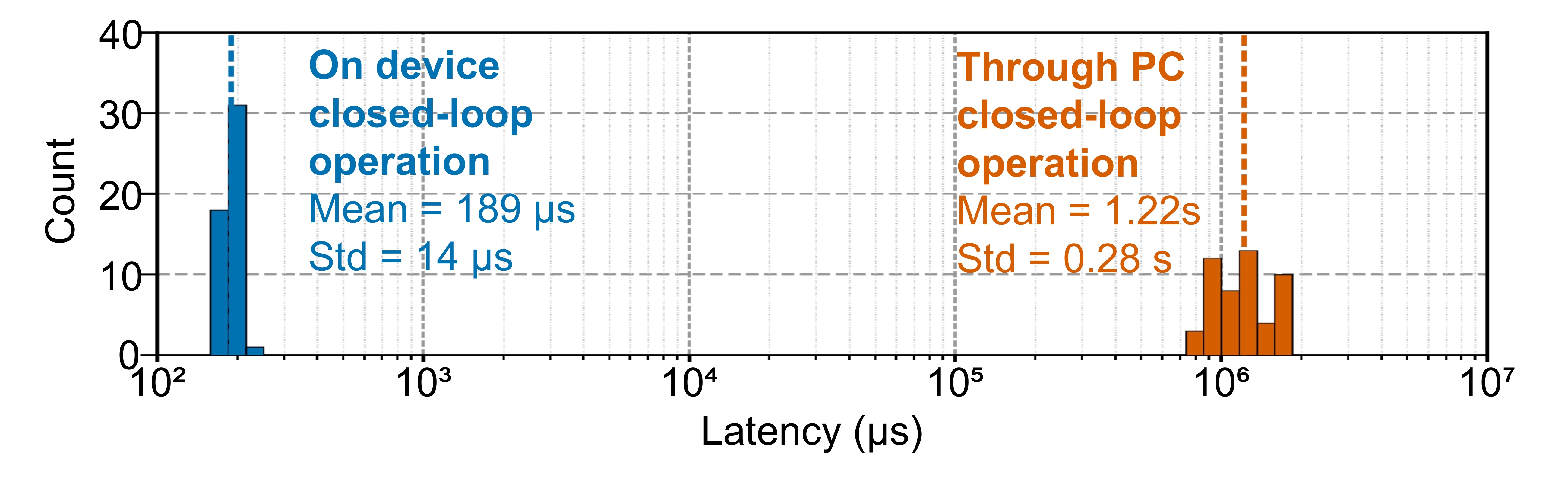} \vspace{-8mm}
    \caption{Latency of spike-triggered stimulation through two pathways, one operating fully on-device and the other completing a round trip through the host PC.}
    \label{fig:exp_CL}
\end{figure}

To achieve spike-triggered stimulation, we implemented a routine in which the M4 sends a stimulation command to the M7 via an HSEM interrupt when a spike is detected. Fig.~\ref{fig:exp_CL} compares the latency of the on-device spike-triggered stimulation path with that of a closed-loop path mediated by the host PC, demonstrating an improvement of three orders of magnitude. This result validates the system's capacity for sub-millisecond triggered stimulation and establishes the benefit of moving closed-loop operations on-device.
\vspace{-5pt}

\vspace{4mm}
\section{Conclusion}\label{conclusion}
This work presents an open-architecture bidirectional neural interface built around the dual-core STM32H747 and commercial Intan chips, enabling on-device AI without sacrificing recording or stimulation channel counts, as shown in Table~I. The dual-core processing scheme decouples acquisition and inference, while the two-stream memory hierarchy supports a range of model architectures.

\begin{table}[!h]
\centering
\label{tab:comparison}
\caption{Comparison of Neural Interface Systems with Off-the-Shelf Front Ends}
\scriptsize
\renewcommand{\arraystretch}{0.95}
\setlength{\tabcolsep}{3pt}
\begin{adjustbox}{max width=\columnwidth}
\begin{tabular}{l|ll|cc|c|cc}
\toprule
\textbf{System} & \textbf{Front-End} & \textbf{Compute} & \textbf{Rec.} & \textbf{Max} & \textbf{Stim.} & \textbf{CL*} & \textbf{ML} \\
 & \textbf{/ Stim} & & \textbf{Ch.} & \textbf{kS/s} & \textbf{Ch.} & & \\
\midrule
\textbf{This Work} & \textbf{RHD+RHS} & \textbf{STM32H747} & \textbf{128} & \textbf{20} & \textbf{64} & \textbf{Yes} & \textbf{Yes} \\
\midrule
\multicolumn{8}{l}{\textit{\textbf{Host-PC Closed-Loop}}} \\
\midrule
M3BA~\cite{LuhmannEtAl}        & ADS1299         & Cortex-M4        & 6 & 16 & N/A & No  & No \\
Trumpis et al.~\cite{Trumpis2017-zx}        & DDC 264         & ICE 40 FPGA        & 61 & 2 & N/A & No  & No \\
Open Ephys~\cite{Siegle_Lopez_Patel_Abramov_Ohayon_Voigts_2017}        & RHD         & XEM6310 FPGA        & 512 & 30 & ext. & No  & No \\
ONIX~\cite{Newman_Zhang_Cuevas-Lopez_Miller_Honda_vanGoes_Leighton_Carvalho_Lopes_Lakunina_etal._2025}             & RHD     & MAX10 FPGA       & 64        & 30       & 1    & No  & No \\
Iris 128B~\cite{Jacobs_Monge_Switalla_Frederick_Deku_2025}        & RHS & N/A  & 128       & 30 & 128  & No  & No \\
\midrule
\multicolumn{8}{l}{\textit{\textbf{On-Device Closed-Loop}}} \\
\midrule
Neurochip3~\cite{shupe2021neurochip3}       & RHD  & Cyclone V FPGA & 32        & 20    & 6    & SNN & No \\
Ferrero et al.~\cite{ferrero2025closed}      & RHD     & STM32          & NR  & NR       & ext. & BP  & No \\
\bottomrule
\end{tabular}
\end{adjustbox}
\par\vspace{5pt}
\raggedright\scriptsize
CL*: On-device closed-loop. NR: Not reported. BP: Band-power thresholding; ext.: External.
\end{table}

\section*{Acknowledgment}
This research was supported in part by the Natural Sciences and Engineering Research Council of Canada (NSERC), the University of Toronto, and University Health Network (UHN). We acknowledge Mansheel Chahal, Zhuofan Yang, and Lin Sheng for their contributions to this project.

\bibliographystyle{IEEEtran}
\bibliography{ref}

\end{document}